# Computational Screening Study towards Redox-Active Metal-Organic Frameworks

Jelena Jelic[1], Dmytro Denysenko[2], Dirk Volkmer[2] and Karsten Reuter[1]

[1] *Technische Universität München, Department Chemie, Lichtenbergstr.4, 85747, Garching, Germany*
[2] *Universität Augsburg, Institute of Physics, Universitätsstrasse 1, 86159, Augsburg, Germany*

jelena.jelic@mytum.de

**Abstract:** The metal-organic framework (MOF) MFU-4*l* containing Co(II) centers and Cl[-] ligands has recently shown promising redox activity. Aiming for further improved MOF catalysts for oxidation processes employing molecular oxygen we present a density-functional theory (DFT) based computational screening approach to identify promising metal center and ligand combinations within the MFU-4*l* structural family. Using the $O_2$ binding energy as a descriptor for the redox property, we show that relative energetic trends in this descriptor can reliably be obtained at the hybrid functional DFT level and using small cluster (scorpionate-type complex) models. Within this efficient computational protocol we screen a range of metal center / ligand combinations and identify several candidate systems that offer more exothermic $O_2$ binding than the original Co/Cl-based MFU-4*l* framework.

## 1. INTRODUCTION

Next to applications in gas storage or drug delivery, metal-organic frameworks (MOFs) receive increasing attention as redox active catalysts [1-9]. Combining bi- or multifunctional ligands (predominantly polycarboxylate ligands) and (transition) metal ions, moderately robust MOFs can be prepared. Advantages of MOFs as heterogeneous catalysts are enhanced catalyst stability due to the spatial separation of single catalytic sites in the framework [10], high porosities in the absence of any non-accessible bulk volume (dead volume) [6] and – most evidently – their pore size(s), and thus their substrate shape and size selectivity, that can be systematically tailored by employing different organic linkers [11]. Three conceptually different approaches have been used to gain catalytically active MOFs: introducing catalytic metal centres at the nodes of the constituting secondary building units (SBU), attaching catalytically active functional groups or coordination units at the MOF struts (i.e. the organic linkers), and embedding nanosized metal clusters within the pores of the MOF. Up till now, the use of catalytically active metal centres positioned at the nodes of the MOF-constituting SBUs was predominantly explored for Lewis acid catalysis [12-15], whereas examples on redox catalysis are rare.

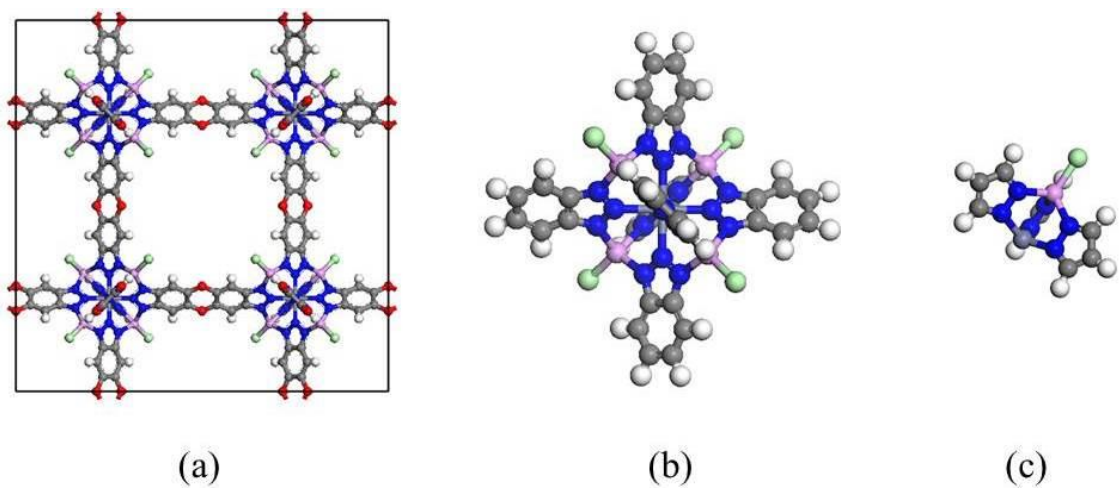

**Figure 1.** Ball-and-stick models of a) MFU-4*l* cubic unit cell , b) secondary building unit (SBU): $[M^{a(II)}M^{b}_{4}X_{4}(L_{6})]$, L = 1,2,3 - triazolate i.e. Kuratowski complex, c) metal-modified "scorpionate" complex (see text). Gray, C; blue, N; red, O; white, H atoms. Pink, $M^{b}$ metal center; dark gray, $M^{a(II)}$ central octahedrally coordinated metal center; green, $X^{-}$ ligand.

In terms of catalytic transformations especially MFU-4-type frameworks show several highly promising features [16]. Particularly large pore sizes were achieved by the use of bis-triazolo-dibenzo-dioxin (BTDD) linkers, which then lead to the so-called MFU-4*l* members within this class of MOFs (figure 1a) [17]. Their Kuratowski-type pentanuclear SBUs [$M^{a(II)}M^b_4X_4(L_6)$] (figure 1b) offer four well accessible, tetrahedral metal centers $M^b$ each being coordinated by three N donor ligands. Most appealing for redox catalysis is that each such coordination site is stereochemically and topologically related to the typical coordination motif found in so-called scorpionates (= metal complexes with hydrotris(pyrazolyl)borato ligands, figure 1c), a famous class of coordination compounds that has been invented and developed by S. Trofimenko since the early 1960's [18]. Dioxygen activation has already been reported for cobalt scorpionates [19], as has the feasibility of oxygen insertion into aliphatic C-H bonds [20]. While cobalt complexes seem thus particularly promising for catalytic oxidation processes [21], the search for other catalytically active metal oxo complexes that can activate and insert oxygen into C-H bonds remains a flourishing area of active research [22].

Since at present there is no workable concept of incorporating scorpionate complexes into porous frameworks, MFU-4-type frameworks comprising this coordination motif could fill this gap, from which finally a novel class of technically and commercially attractive heterogeneous catalysts may evolve. Recently, a first step along this route was achieved by the demonstrated redox activity of a Co-based MFU-4*l* framework [23]. Furthermore, this study revealed an important additional advantage of MFU-4*l* frameworks in terms of the possibility for a post-synthetic metal center substitution, i.e. in principle there is a generic route to also explore other redox active metal centers. Notwithstanding, the iterative and time-consuming development of efficient syntheses leading to phase-pure MOFs, as well as their structural characterization imposes severe limitations on the number of compounds which can be examined in catalytic test reactions. This is particularly vexing, when considering that the non-radical catalytic oxidation or oxygenation activity will also critically depend on the interplay between the metal center and the electronic nature of the ligand $X^-$, cf. Fig. 1, i.e. there is a quite large matrix of MOFs that needs to be explored.

In this situation, computational screening can step in as most valuable tool to identify interesting ligand-metal center combinations on which ensuing synthesis activities would then be focused. An intuitive and well computable descriptor for the redox properties is in this context simply the $O_2$ binding energy to the metal center, reflecting the degree of bond activation. Such a screening study is only of use though, if it is much faster than experimental synthesis, yet still sufficiently predictive. The prior demand necessitates the use of numerically efficient approaches. While we will show below that highest efficiency can be reliably achieved with respect to system size by focusing the calculations on the small scorpionate-type moieties, the situation is less clear with respect to the level of theory necessary to reach predictive quality. The intricate electronic structure of metal-oxo complexes is generally known to severely challenge the present-day workhorse in terms of efficient electronic structure calculations, density-functional theory (DFT) with semi-local or hybrid functionals [24, 25]. This is most clearly exemplified for the present case by the qualitative differences reported recently for the oxygen binding energy in a similar Co-based MOF within the MFU-1 structural family [26]. At the gradient-corrected level this binding was obtained as exothermic, while at the hybrid level it was endothermic with the absolute difference between the two computed binding energies exceeding 1 eV!

In this work we address this problem by tracing it largely back to the self-interaction induced deficiency of the semi-local functional in properly describing the change of metal spin state concomitant to $O_2$ adsorption. As this is rather independent of the actual ligand, screening calculations at corresponding levels of DFT theory are still meaningful as they only require relative energetic trends to identify suitable MOF candidates, not the absolute binding energy itself. Demonstrating the additional possibility to perform these calculations on the small scorpionate-type units, this brings us into a position to efficiently screen a range of metal centers and ligands. The results clearly identify a small number of interesting combinations, among which particularly Mn and Fe centers with $NH_2^{-1}$ ligands additionally prevent possible limitations due to spin-transition induced $O_2$ adsorption barriers.

## 2. THEORY
All DFT calculations have been performed with the all-electron full-potential code FHI-aims [27,28]. Its localized basis sets based on numeric atom-centered orbitals (NAO) allow to perform the calculations both for finite molecular systems and infinite systems through the use of periodic boundary conditions (in the present case with Γ-point sampling). Electronic exchange and correlation was treated on the level of the generalized gradient approximation (GGA) PBE functional [29] or on the level of the hybrid B3LYP functional [30,31]. Dispersive interactions lacking at these levels of theory were considered through the dispersion-correction scheme due to Tkatchenko and Scheffler [32], and we correspondingly denote this as PBE+TS and B3LYP+TS below. The central quantity used as descriptor in the computational screening approach is the $O_2$ binding energy defined as

$$E_b(O_2) = [\, E_{tot}(O_2@MOF) - E_{tot}(MOF) - E_{tot}(O_2) \,] \;,$$

where $E_{tot}(O_2@MOF)$, $E_{tot}(MOF)$, and $E_{tot}(O_2)$ are the total energies computed for $O_2$ adsorbed at the MOF, for the clean MOF, and for the isolated $O_2$ gas-phase molecule, respectively. The calculation of this numerically inexpensive quantity is sufficient for the here required relative energetic trends, and could ultimately be augmented by vibrational calculations to allow comparison to experimentally measured reaction enthalpies. In the employed sign convention negative binding energies indicate exothermicity. Geometry optimization was generally performed at the PBE+TS level until residual forces fell below $10^{-4}$ eV/Å. Hybrid B3LYP+TS energetics was subsequently obtained through single-point calculations on these optimized geometries. Systematic convergence tests show $E_b(O_2)$ to be sufficiently converged for the intended relative energetic trend study (i.e. within ±50 meV) at the "tier2" NAO basis set and when using tight integration grids. Spin and charge assignments in the text are based on calculated Hirshfeld projections.

## 3. RESULTS

### 3.1 Definition of screening space: ligand and metal centers

The starting point for the present study is the reported redox activity of the peripheral Co(II) centers contained within the coordination units of a MFU-4$l$ [$M^a Co^b_4 Cl_4 (ta)_6$] ($M^a$ = Zn or Co, , tatriazolate ligand) framework [23]. Preserving the structural character of the MFU-4$l$ class of MOFs, obvious routes to be explored towards a further improved activity would be metal centers other than Co and ligands other than $Cl^-$. In terms of ligand substitution we hereby concentrate on ligands that keep the metal center in the same formal charge state as for $Cl^-$, namely $OH^-$, $H^-$, $CN^-$, $F^-$, $NH_2^-$, $NO_2^-$, $NO_3^-$, and $CF_3SO_3^-$. These anions have been chosen because they are representative of ligands comprising different sigma and pi-metal bonding capabilities, i.e. they cover the whole range of different ligand types described by the so-called spectroscopic series of ligands found in classical monographs of ligand-field theory (*vide infra*). Among these $OH^-$, $Cl^-$, $NO_3^-$ and $CF_3SO_3^-$ are considered as weak-field ligands providing different denticities ($OH^-$, $Cl^-$, monodentate; $NO_3^-$, bidentate; $CF_3SO_3^-$, bi- or tridentate), whereas $F^-$, $NO_2^-$, and $CN^-$ were chosen owing to their well-known propensity to interact via additional metal-ligand π-bonds and thus are regarded as strong-field ligands. $NH_2^-$ and $H^-$, finally, serve as strongly basic and reducing ligands. As to the metal centers significant reactivity with C-H bonds has particularly been observed for metal-oxo complexes that are significantly destabilized by occupied $d$-orbitals (i.e., possessing $d^n$ configurations with $n$ = 4 or higher) [22]. Only then is the metal-oxo multiple bond sufficiently weakened to render alternatives, such as the formation of hydroxo groups, energetically competitive. Aiming to produce M=O units from M-$O_2$ precursors, we correspondingly focus within the 3$d$ transition metal series particularly on Mn, Fe, Co and Ni centers.

For the metal-dioxygen interaction we can generally expect the existence of a weak physisorptive adsorption mode. In such a van der Waals bonded complex the dioxygen molecule will preserve its gas phase spin triplet state. In addition to this, direct coordination to a single metal center can occur in a side-on $\eta^2$-mode or an end-on $\eta^1$ mode. We systematically tested both chemisorptive adsorption modes for the Co-based scorpionates described below. In full agreement with previous DFT work on the similar Co-based MFU-1 framework [26], the side-on $\eta^2$-mode in which the peripheral Co(II) centers of MFU-4$l$ effectively switch to an octahedral configuration results as by far most favourable for all monodentate ligands (> 0.4 eV, except for the $NH_2^{-1}$ ligand where the difference is only 0.15 eV). For bidentate ligands ($NO_2^-$, $NO_3^-$, $CF_3SO_3^-$) this $O_2$ $\eta^2$-binding mode implies a switching to monodentate ligand coordination, but we found this to be still more favourable or at least energetically en par to the alternative end-on $\eta^1$ $O_2$ coordination with bidentate ligand. All screening calculations are correspondingly restricted to the side-on $\eta^2$ chemisorptive mode.

### 3.2 Reduction of system size

The primitive trigonal cell derived from the original face-centered cubic unit cell of MFU-4$l$ with a simple

$Cl^-$ ligand comprises 162 atoms. While the localized basis sets within FHI-aims allow to efficiently treat such infinite systems within a periodic boundary condition setup, the computational costs are still noticeable, in particular at the hybrid functional level. Suspecting the $O_2$ binding to the tetrahedral metal centers to be predominantly determined locally, we therefore additionally consider two types of finite cluster models built from the MOF SBU. On the one hand this is the Kuratowski complex shown in Fig. 1b, which is obtained by truncating the linkers of one entire SBU after the phenyl moiety and saturating the broken bonds with hydrogen (89 atoms). On the other hand this is the complex shown in Fig. 1c (and in a different perspective in Fig. 2), which only contains a single tetrahedral active center and three adjacent pyrazole units (27 atoms). This closely mimics the coordination environment of one of the tetrahedral metal centers of the Kuratowski complex, but could similarly be viewed as a metal-modified scorpionate complex, in which the HB unit of a typical scorpionate is replaxed by a Zn(II) metal ion. For simplicity we will refer to this complex as scorpionate complex in the following, even though it is formally rather a „metal-modified" scorpionate.

**Table 1.** PBE+TS $O_2$ binding energy (eV) in the $O_2^{-1}$ side-on binding mode at two cluster models to Co-MFU-4*l* and for several ligands (see text).

| $E_b(O_2)$ PBE+TS | $Cl^-$ | $OH^-$ | $H^-$ | $CN^-$ | $F^-$ | $NH_2^-$ |
|---|---|---|---|---|---|---|
| Scorpionate | -0.68 | -1.09 | -1.97 | -1.19 | -0.92 | -1.24 |
| Kuratowski | -0.58 | -1.02 | -1.99 | -1.08 | -0.84 | -1.26 |

Table 1 summarizes the calculated PBE+TS $O_2$ binding energy at the two cluster models for a Co(II) center and a series of ligands. For the $O_2^{-1}$ side-on binding mode already the difference between the two cluster models is only of the order of ~0.1eV, and the local bonding geometries around the metal center are essentially identical. Equivalent calculations for physisorbed $O_2$, i.e. the van der Waals bonded complex, show even smaller energetic differences of the order of 40 meV. For $Cl^-$ and $CN^-$ ligands both adsorption modes have also been computed in the fully periodic MFU-4*l* framework, revealing insignificant energetic and geometric differences to the results obtained for the Kuratowski complex (< 20 meV). Finally, $O_2^{-1}$ side-on binding energies at scorpionate and Kuratowski complexes were also calculated at the B3LYP+TS level, and exhibited differences of the same order of magnitude as those found at the PBE+TS level. We correspondingly conclude that the oxygen binding at the smallest scorpionate complex mimics the one in extended MFU-4*l* quite faithfully, with the indicated uncertainty of ~0.1 eV in the binding energies not relevant for the intended screening approach. All calculations reported from now on are therefore conducted most efficiently for this small cluster model.

3.3 *Energetic trends at gradient-corrected and hybrid DFT level*

**Table 2**. $O_2^{-1}$ side-on binding energy (eV) at Co-scorpionate for several ligands. Shown are results at PBE+TS and B3LYP+TS level, as well as the difference between them.

| $E_b(O_2)$ | $Cl^-$ | $OH^-$ | $H^-$ | $CN^-$ | $F^-$ | $NH_2^-$ | $NO_2^-$ | $NO_3^-$ | $CF_3SO_3^-$ |
|---|---|---|---|---|---|---|---|---|---|
| PBE+TS | -0.68 | -1.09 | -1.97 | -1.19 | +0.92 | -1.24 | -1.01 | -1.01 | -0.75 |
| B3LYP+TS | +0.38 | -0.02 | -0.73 | -0.10 | +0.10 | -0.18 | +0.20 | +0.03 | +0.25 |
| Difference | **-1.06** | **-1.07** | **-1.24** | **-1.09** | **-1.02** | **-1.06** | **-1.21** | **-1.04** | **-1.00** |

As mentioned in the introduction the conflicting results for gradient-corrected and hybrid DFT reported by Tonigold *et al.* for the highly similar MFU-1 system [26] question the reliability of either level of theory for the description of the $O_2$ – MOF interaction. Before embarking on the actual screening study we therefore further analyse this difference for the present system. Within the perspective of a screening approach that primarily intends to identify energetic trends over a series of possible ligands or metal centers, we thereby concentrate on Co based scorpionates and compute the $O_2$ binding energy for a range of ligands. The results summarized in table 2 reveal a similar disagreement between gradient-corrected and hybrid DFT as observed for the MFU-1 system [26], with PBE+TS yielding a significantly stronger binding. Intriguingly, however, this difference of about 1.1 eV is rather independent of the various ligands, which suggests that its origin lies primarily in the treatment of the Co metal center itself.

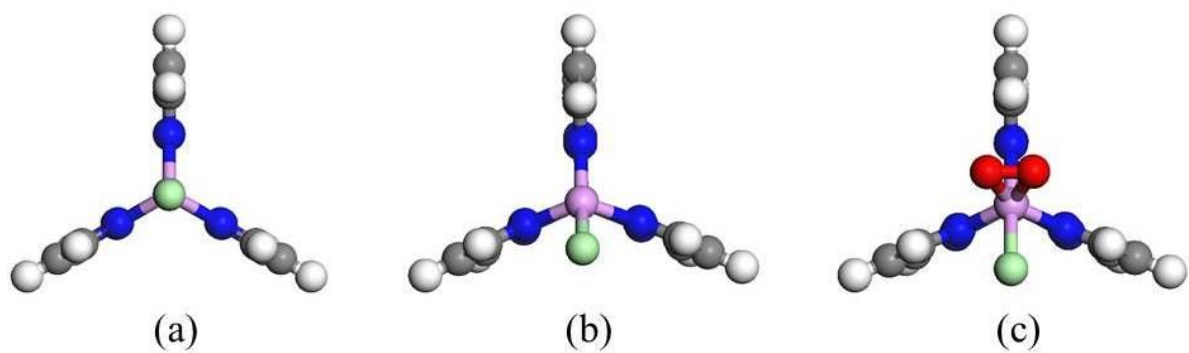

**Figure 2.** Optimized Co scorpionate geometries for a $Cl^-$ ligand: a) plain structure – high-spin; b) plain structure – low-spin; c) $O_2^{-1}$ side-on binding mode. The color code used for the different species is the same as in Fig. 1.

Thinking along this line we recall that $O_2$ binding in the side-on mode effectively changes the Co coordination from tetrahedral to octahedral. Figure 2 exemplifies the corresponding optimized geometries for the $Cl^-$ ligand, with equivalent results obtained for all other ligands tested. In agreement with the expectations from ligand-field theory this coordination change is accompanied with a change in the Co spin state, cf. figure 3: In the tetrahedral coordination of the plain MOF both functionals predict a quartet high-spin state for the Co(II) center, whereas a singlet low-spin state is predicted for the octahedrally coordinated Co(III) after dioxygen binding. Suspecting the concomitant $d$-orbital rehybridization to largely contribute to the discrepancy between the two levels of DFT theory, we perform calculations in which we already force the plain MOF structure to the low-spin state, namely a doublet state for Co(II). As apparent from figure 2 this immediately breaks the tetrahedral symmetry, with the $Cl^-$ ligand moving towards the position it also exhibits after dioxygen adsorption.

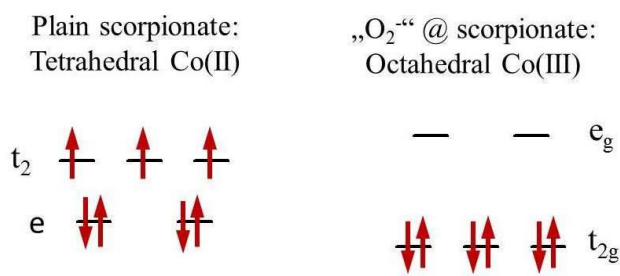

**Figure 3.** Electron arrangement in Co(II) and Co(III) $d$-orbitals in tetrahedral and octahedral coordination according to ligand-field theory.

Table 3 compiles the energetic difference between this low-spin state and the true high-spin state as obtained by PBE+TS and B3LYP+TS. Interestingly, we again find a rather large and systematic difference between these two levels of theory. For all tested ligands, the hybrid functional yields a much larger energetic gap to the low-spin state, i.e. the rehybridization concomitant with $O_2$ side-on adsorption is energetically much more costly than at the gradient-corrected level. If we correspondingly compute the $O_2$ binding energies with respect to the hypothetical low-spin plain scorpionate, the discrepancy between the two DFT descriptions is much reduced, namely by exactly the difference in the spin preference shown in table 3. While for instance for the $Cl^-$ ligand the real $O_2$ binding energy computed with PBE+TS is by 1.06 eV more exothermic than for B3LYP+TS, cf. table 2, this is reduced to only (1.06 eV - 0.46 eV) = 0.6 eV with respect to the hypothetical low-spin scorpionate, cf. table 3.

**Table 3**. Energy difference (eV) between the high- and low-spin state of a Co-scorpionate with various ligands (see text). Shown are results for PBE+TS and B3LYP+TS, as well as the difference between them.

| Spin preference | $Cl^-$ | $OH^-$ | $H^-$ | $CN^-$ | $F^-$ | $NH_2^-$ | $NO_2^-$ | $NO_3^-$ | $CF_3SO_3^-$ |
|---|---|---|---|---|---|---|---|---|---|
| PBE+TS | -0.50 | -0.28 | -0.10 | -0.23 | -0.43 | -0.13 | +0.16 | +0.17 | -0.11 |
| B3LYP+TS | -0.96 | -0.78 | -0.72 | -0.77 | -0.91 | -0.65 | -0.36 | -0.36 | -0.49 |
| Δ (PBE-B3LYP) | 0.46 | 0.50 | 0.62 | 0.54 | 0.48 | 0.52 | 0.52 | 0.53 | 0.38 |

Our analysis thus traces a large part of the discomforting discrepancy between the two DFT functionals back to the different ability of the gradient-corrected and hybrid functional in describing the high-spin state of the tetrahedrally coordinated Co(II) center in the plain MFU-4*l* structure. Specifically, we suspect the generally acknowledged electron delocalization problem of the gradient-corrected functional [33] to lead to a wrong account of the fully-occupied $d_e$ orbitals. This interpretation gets support from the observation that in the calculated PBE+TS density of states (DOS) these orbitals are partly found at higher energies than the $d_{t2}$ orbitals, in contrast to the predictions from ligand-field theory, cf. figure 3. In contrast, in the B3LYP+TS DOS this is no longer the case, i.e. here all $d_e$ orbitals are consistently located below the $d_{t2}$ orbitals. In fact, we can further corroborate the central relevance of this $d$ orbital ordering by artificially enforcing the correct ordering at the PBE+TS level through the application of an appropriate U operator [34]. In corresponding PBE+U+TS calculations, in which the U value of 1.2 eV has been set to fix the $d_e$ orbitals at the position they have in the B3LYP+TS calculations, we then indeed obtain $O_2$ binding energies that are very close to those obtained with the latter functional (within 0.1 eV).

We therefore conclude that the discrepancy between the gradient-corrected and hybrid functional calculations are predominantly due to the wrong $d$ orbital positioning predicted by the prior electron-delocalization riddled functional. As this is mostly independent of the actual ligand employed, already calculations at the computationally most efficient PBE+TS level should yield reliable energetic trends within the space of screened ligands. On the other hand, we cannot expect the same to hold for different metal centers, i.e. electron delocalization induced errors in the $d$ orbital positions will generally be different for different metals. This view is confirmed by evaluating the difference in computed $O_2$ binding energies at PBE+TS and B3LYP+TS level for scorpionates with Mn, Fe and Ni centers. Consistent with the results presented for Co based scorpionates in table 2 we each time find the energetic difference between the two levels of theory to be rather independent of the actual ligands. Notwithstanding, this difference varies largely for the different metal centers. Whereas it was ~1.1 eV for the above discussed case of Co, it is ~0.55 eV, ~0.25 eV and ~0.9 eV for Mn, Fe and Ni, respectively. Such variations preclude a comparison of the $O_2$ binding energy descriptor for different metals at the PBE+TS level. We correspondingly perform the actual screening calculations over the full range of metals and ligands at the B3LYP+TS level. At this level the electron delocalization error is largely removed through the admixture of exact exchange [35] and we expect reliable energetic trends also across the different metal centers tested.

3.4 *Screening the ligand/metal center matrix*
Within the established computational screening approach (scorpionate cluster model, DFT B3LYP+TS) we proceed to compute the $O_2$ binding energy in both the van der Waals physisorptive mode and the side-on chemisorptive mode for a range of ligands and different metal centers. As expected, physisorption is possible at all tested metal centers. The bond strength varies little with the actual ligand and metal center, and lies between -0.1 and -0.2 eV throughout. Suspecting that such a weak binding does not allow for sufficient bond activation, we thus concentrate on the chemisorptive mode and aim to identify metal center-ligand combinations that yield more exothermic binding energies than those offered by physisorption. The results are compiled in table 4 and reveal several interesting trends. First of all, the $Cl^-$ ligand employed in the original experiments for Zn and Co based MFU-4*l* [23] yields generally a very weak binding, and concomitantly low bond activation. From the calculations in the hypothetical low-spin state for the plain scorpionate we attribute this to a rather high energy cost for the spin change and rehybridization to octahedral coordination in the presence of this ligand. This rehybridization cost is generally also the reason for the quite large number of in fact endohedral binding energies computed: The actual $O_2$-MOF interaction is certainly stronger than in the physisorptive mode and we validated that in all cases the chemisorptive binding mode is metastable, i.e. corresponds to a minimum on the potential energy surface. Nevertheless this energetic gain is counterbalanced by the rehybridization cost to octahedral coordination, which in many cases leads to a net endothermic binding energy.

Table 4. DFT B3LYP+TS $O_2$ binding energies (eV) at scorpionates with different metal centers and ligands. Interesting candidates with exothermic binding energies below -0.3 eV are marked in bold.

| $E_b(O_2)$ B3LYP+TS | $Cl^-$ | $OH^-$ | $H^-$ | $CN^-$ | $F^-$ | $NH_2^-$ | $NO_2^-$ | $NO_3^-$ | $CF_3SO_3^-$ |
|---|---|---|---|---|---|---|---|---|---|
| Mn | +0.25 | **-0.43** | **-0.37** | -0.09 | -0.11 | **-0.60** | +0.17 | +0.25 | +0.27 |
| Fe | +0.03 | **-0.32** | -0.26 | +0.03 | -0.24 | **-0.41** | +0.01 | +0.01 | +0.08 |
| Co | +0.38 | -0.02 | **-0.73** | -0.10 | +0.10 | -0.18 | +0.20 | +0.03 | +0.25 |
| Ni | +0.77 | +0.29 | **-0.42** | +0.38 | +0.50 | +0.06 | -0.13 | +0.73 | +0.92 |

Encouragingly, other ligands than the original $Cl^-$ seem to allow for more flexibility, cf. table 3, and correspondingly show more exothermic $O_2$ binding. Particularly interesting in this respect are the $H^-$, $OH^-$, and $NH_2^-$ ligands. There is, however, a non-trivial correlation between ligand and metal center. A ligand that shows high exothermicity for one metal center does not necessarily show this for another metal center, too. In other words, the energetic trend across the tested ligands is different for the different metal centers. This dictates the explicit computation of the full ligand – metal center matrix, rather than a selective optimization among the ligands for one metal center and subsequent optimization of the metal center for this optimized ligand.

Distributed over the matrix we find several metal center-ligand combinations that yield a binding energy more exothermic than -0.3 eV, and thus promise a bond activation significantly stronger than in the purely physisorptive adsorption mode. Before concluding on these combinations as interesting candidates for experimental synthesis, one has to recognize though that computational screening is only as useful as the actual descriptor employed. The computed binding energy is an intuitive measure of bond activation and ensuing redox activity. Yet, it does not tell about unwanted side reactions that e.g. compromise the structural stability of the entire MOF. In this respect, the use of $H^-$ as ligand could be critical, since we can expect a further reaction of bound $O_2$ molecules leading to hydroperoxide species, which cannot be transformed back to hydride and thus no catalytic activation of $O_2$ can be achieved.

Furthermore, a computed thermodynamic metastability of the chemisorptive binding mode does not tell about possible kinetic limitations in reaching this state. Here, we particularly note the largely differing spin properties of the Co centers compared to the other metal centers tested. Consistent with the expectations from ligand-field theory we find the change from tetrahedral to octahedral coordination concomitant with $O_2^{-1}$ side-on binding to be accompanied with a high-spin to low-spin transition in the case of cobalt. Specifically and independent of the actual ligand, the Co(II) center in the plain framework changes from its quartet spin state to a singlet state as Co(III) in the octahedral configuration after $O_2^{-1}$ binding. Together with the spin transition of the $O_2$ molecule from its triplet gas-phase state to the doublet state after adsorption, this amounts to a total change in spin multiplicity of $\Delta S=4$. In contrast, for Mn and Ni we obtain smaller spin multiplicity changes of $\Delta S=2$, and for Fe a $\Delta S=0$, since Mn and Fe metal centers stay in their high-spin state after adsorption. In ligand field theory such a high multiplicity change in the case of Co necessarily imposes a pairing energy penalty on the reaction, since two electrons formerly in singly occupied orbitals need to be paired into one orbital, cf. Figure 3. If this happens before the energy gain due to the $O_2$ chemisorption sets in a barrier results along the reaction path even in an electronically adiabatic picture, where additional spin selection rules are not even considered [36-38]. As illustrated in figure 4 for a $NH_2^{-1}$ ligand this is indeed what we find in our adiabatic DFT calculations, i.e. we obtain a sizable activation barrier of the order of 0.6 eV to reach the chemisorptive state. Equivalent calculations for other ligands confirm that such a high activation barrier is a specific property of the Co centers, whereas in particular for the promising combination Fe or Mn center with $NH_2^{-1}$ ligand we obtain vanishingly small activation barriers. In light of these findings these two systems appear finally as most promising candidates for experimental synthesis among those included in our screening study: both yield exothermic binding energies in the side-on mode that are significantly stronger than pure physisorption, cf. table 4, and this adsorption mode can be reached without kinetic limitations.

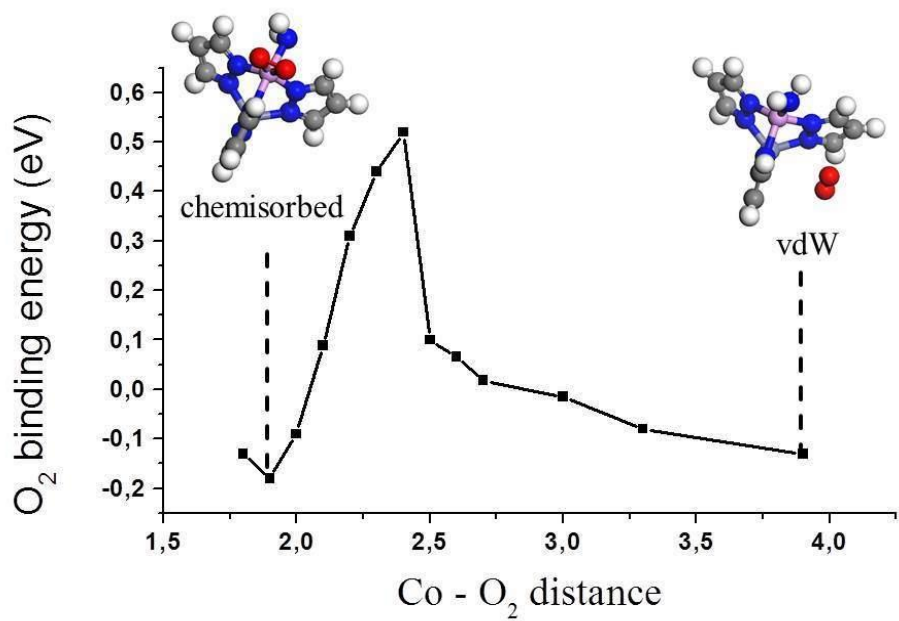

**Figure 4.** Calculated B3LYP+TS energy profile between the vdW physisorptive and $O_2^{-1}$ side-on chemisorptive binding mode at a Co-scorpionate with $NH_2^-$ ligand and using the $O_2$-metal center distance as reaction coordinate.

## 4. CONCLUSIONS

We have performed a computational screening study to identify promising alternative metal center – ligand combinations for the MFU-4*l* metal-organic framework that could improve on the redox activity achieved recently with Co metal centers and $Cl^-$ ligands [23]. Using the $O_2$ binding energy as descriptor to assess this redox activity, we first established that calculations at the DFT hybrid functional level and using small Scorpionate-type cluster models represent a numerically efficient and reliable computational protocol for the required relative energetic trends. Screening Mn, Fe, Co, and Ni centers together with a range of ligands we identify several candidates that afford a chemisorptive $O_2^-$ side-on binding mode that is notably more stable than mere van der Waals physisorption. Allowing to reach this chemisorptive state without significant activation barriers, particularly Fe or Mn centers with $NH_2^{-1}$ ligands appear as most promising for an ensuing experimental synthesis. A major limiting factor towards even larger $O_2$ activation is the high energy cost connected with the rehybridization from tetrahedral to octahedral coordination concomitant to side-on $O_2^{-1}$ adsorption. This suggests to extend the screening approach either to metal(II) centers with higher reactivity towards oxygen (e.g. $V^{II}$, $Cr^{II}$) or to ligands that additionally stabilize the coordinated dipolar dioxygen e.g. through intramolecular hydrogen addition.

We gratefully acknowledge funding with the priority program 1362 "Porous Metal-Organic Frameworks (MOFs)" of the Deutsche Forschungsgemeinschaft (DFG).